\documentclass[twocolumn]{aastex631}
\vspace{-7mm}

\usepackage{amsmath, bm}
\newcommand{\tr}{\scalebox{0.6}{$\top$}}
\newcommand{\plus}{\scalebox{0.6}{$+$}}

\vspace{-10mm}
\submitjournal{ApJ}

\begin{document}

\title{The Active Optics System on the Vera C. Rubin Observatory: Optimal Control of Degeneracy Among the Large Number of Degrees of Freedom}

\author[0000-0001-6013-1131]{Guillem Megias Homar}
\affiliation{Department of Aeronautics and Astronautics, Stanford University, Stanford, CA 94305, USA}
\affiliation{Kavli Institute for Particle Astrophysics and Cosmology, Stanford University, Stanford, CA 94309, USA}
\affiliation{SLAC National Accelerator Laboratory, Menlo Park, CA 94025, USA}

\author[0000-0003-4833-9137]{Steven M. Kahn}
\affiliation{Departments of Physics and Astronomy, University of California, Berkeley, Berkeley, CA 94720, USA}
\affiliation{Kavli Institute for Particle Astrophysics and Cosmology, Stanford University, Stanford, CA 94309, USA}
\affiliation{SLAC National Accelerator Laboratory, Menlo Park, CA 94025, USA}

\author[0000-0002-2308-4230]{Joshua M. Meyers}
\affiliation{Kavli Institute for Particle Astrophysics and Cosmology, Stanford University, Stanford, CA 94309, USA}
\affiliation{SLAC National Accelerator Laboratory, Menlo Park, CA 94025, USA}

\author[0000-0002-2495-3514]{John Franklin Crenshaw}
\affiliation{Department of Physics, University of Washington, Seattle, WA 98195, USA}
\affiliation{DIRAC Institute, University of Washington, Seattle, WA 98195, USA}

\author[0000-0002-2308-4230]{Sandrine Thomas}
\affiliation{Vera C. Rubin Observatory Project Office, 950 N. Cherry Ave, Tucson, AZ 85719, USA}

\correspondingauthor{Guillem Megias Homar}
\email{gmegias@stanford.edu}
\begin{abstract}

The Vera C. Rubin Observatory is a unique facility for survey astronomy that will soon be commissioned and begin operations. Crucial to many of its scientific goals is the achievement of sustained high image quality, limited only by the seeing at the site. This will be maintained through an Active Optics System (AOS) that controls optical element misalignments and corrects mirror figure error to minimize aberrations caused by both thermal and gravitational effects. However, the large number of adjustment degrees of freedom available on the Rubin Observatory introduces a range of degeneracies, including many that are \textit{noise-induced} due to imperfect measurement of the wavefront errors. We present a structured methodology for identifying these degeneracies through an analysis of image noise level. We also present a novel scaling strategy based on Truncated Singular Value Decomposition (TSVD) that mitigates the degeneracy, and optimally distributes the adjustment over the available degrees of freedom. Our approach ensures the attainment of optimal image quality, while avoiding excursions around the noise-induced subspace of degeneracies, marking a significant improvement over the previous techniques adopted for Rubin, which were based on an Optimal Integral Controller (OIC). This new approach is likely to also yield significant benefits for all telescopes that incorporate large numbers of degrees of freedom of adjustment.
\end{abstract}



\section{Introduction} \label{sec:intro}

Wide-field telescopes face an arduous challenge: consistently maintaining a high image quality across their field of view. Because they minimize the three main optical aberrations, three-mirror anastigmat telescopes are the most common choice for wide-field surveys. However, they remain vulnerable to aberrations caused by misalignments and mirror deformations induced by gravitational and thermal effects.

The forefront ground-based wide-field survey facility, the Vera C. Rubin Observatory, will soon begin on-sky commissioning on Cerro Pach\'on in Chile. It is poised to address a variety of pressing scientific questions over its 10-year Legacy Survey of Space and Time (LSST) \citep{Ivezic_2019}. The Rubin Observatory incorporates the Simonyi Survey Telescope, an 8.4m-diameter modified Paul-Baker three-mirror anastigmat, with a remarkable $3.5^\circ$ field of view (FOV). To deliver the required seeing-limited image resolution across its field, the tolerance on image aberrations introduced by the telescope and camera is limited to 0.4 arcsec \citep{Claver_2023}. This is planned to be achieved through an Active Optics System (AOS) that uses wavefront sensing to control a large-dimensional parameter space of alignment and figure adjustments on each of the three mirrors, as well as the translation and rotation of the camera \citep{Neill2014}.
 
Historically, wide-field telescopes have allowed for minimizing the gravitational and thermal-led aberrations through the rigid body motions of the different optical elements \textemdash that is, tip, tilt, defocus, and decentering\textemdash and the mirror figure corrections of the primary mirror. Some of these misalignments in three-mirror anastigmats are known to be degenerate when considering only coma, astigmatism and spherical aberration, creating what \cite{Schechter_2011} named a \textit{subspace of benign misalignments}. This degeneracy can be broken when fifth-order moments are measured, as is the case for the Rubin Observatory. However, unlike its predecessors, the Rubin Observatory's large number of degrees of freedom, which include twenty bending modes in the primary/tertiary mirror and twenty in the secondary, introduces a \textit{larger subspace of noise-induced degeneracies} caused by imperfect measurement of the wavefront errors.
  
In the presence of noise in wavefront measurements, the optimal correction for optical aberrations is not uniquely determined. This issue has received much less attention in the literature. Some approaches choose to ignore particular degrees of freedom that are known to create degeneracies. However, that can lead to other problems such as overstressing certain degrees of freedom. Here we introduce a formal approach to addressing this challenge and breaking the degeneracy. We propose the use of a physically-informed characteristic mode basis that optimally removes the linear combinations of degrees of freedom that are degenerate.

Following a review of prior work on wide-field telescopes and a detailed exposition of Rubin Observatory's AOS, we present the mathematical basis behind our proposed solution. We begin by introducing the estimation of the optical state of the telescope from wavefront deviation\textemdash defined as the difference between the estimated wavefront error and the inherent, reference wavefront\textemdash expressed in Zernike coefficients, which incorporates a Singular Value Decomposition (SVD) of the sensitivity matrix. Next, we show how the estimation error is bounded by the noise and the smaller singular values, highlighting the appearance of a noise-induced set of degeneracies. These findings emphasize the importance of operating in a non-degenerate characteristic mode basis. 

For the Rubin Observatory, we exploit detailed image simulations to investigate the impact of noise in the Zernike estimates of the wavefront errors. By employing a power spectrum analysis, we determine the singular value threshold at which noise surpasses signal, thus identifying a distinct subspace of degenerate modes that must be suppressed. Our formulation yields an effective response matrix that quantifies how accurately the algorithm can estimate the true state when operating in this reduced basis. 

Due to the truncation of the SVD basis set, however, corrections associated with perturbations in a particular subset of degrees of freedom can be allocated to a much larger set of actuations in the derived solution. Because of the inherent disparities in the nature of the degrees of freedom, this can lead to pathologies if not properly controlled. We introduce a novel ``weighting approach'' that rescales the sensitivity matrix based on actuator ranges, system flexure, and wavefront influence ranges for each degree of freedom. Through this rescaling, we can strategically incorporate the coupling between different degrees of freedom, emphasizing hexapod motions and secondary mirror corrections over adjustments of the primary/tertiary mirror, which is much more stiff. Next, we elucidate some of the identified characteristic modes and their connection to optical aberrations.

Finally, we demonstrate that applying our improved optical state estimate via a Proportional-Integral-Derivative (PID) control loop ensures convergence to optimal image quality, while effectively avoiding excursions around the noise-induced subspace of degeneracies. In comparison, our approach performs better than Rubin's previous closed-loop Optimal Integral Controller (OIC) control approach, which falls short of overcoming these degenerate movements. We wrap up our discussion by discussing on-sky operational approaches, suggesting enhancements that could benefit the Rubin Observatory AOS, and addressing the implications of our findings not only for the Rubin Observatory but also for the next generation of other major telescope facilities that also incorporate large numbers of degrees of freedom.

\begin{figure*}[t!]
    \centering
    \figurenum{1}
    \plotone{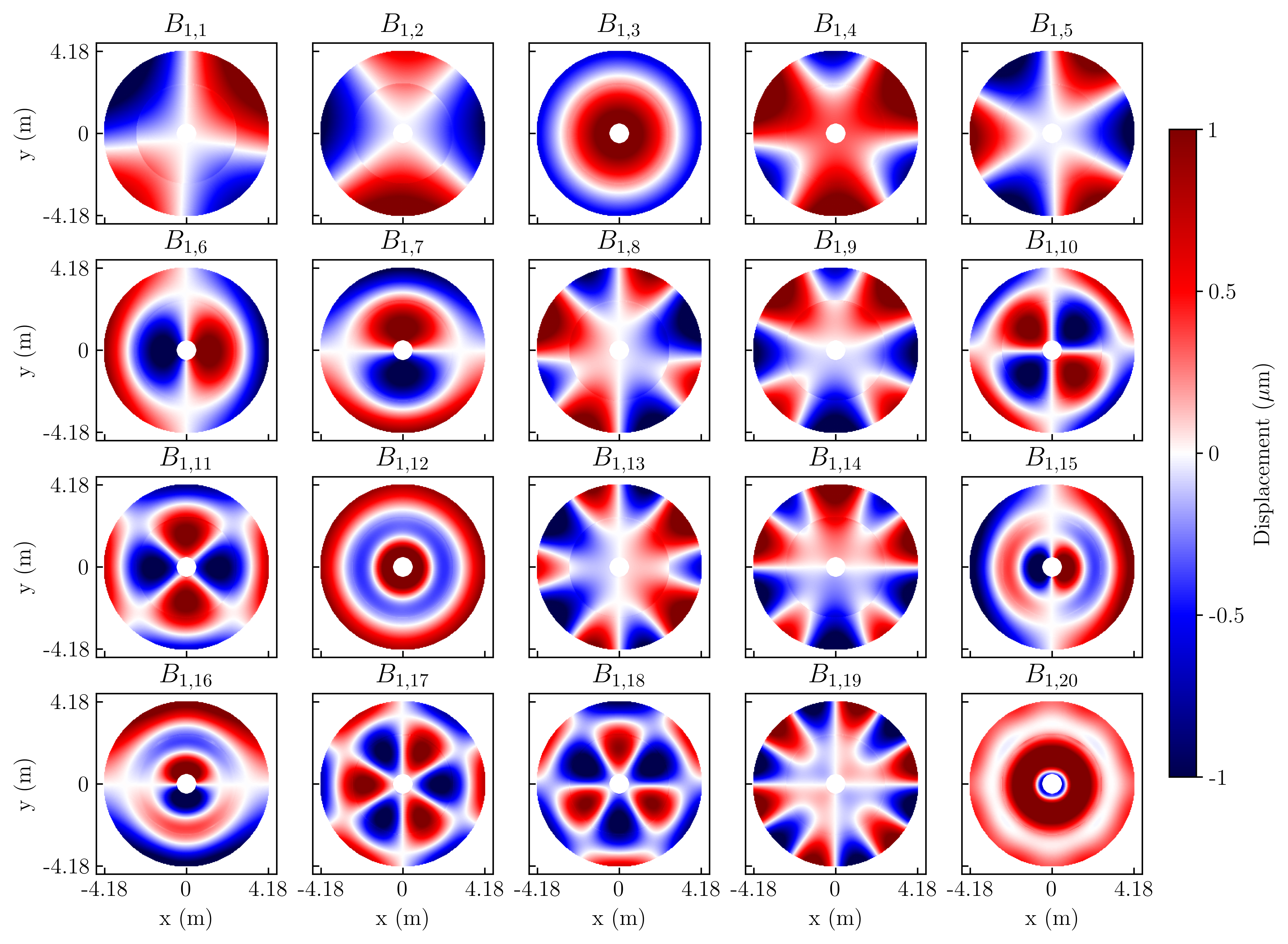}
    \caption{The twenty M1M3 bending modes used in the Rubin Observatory's AOS, arranged sequentially from left to right and top to bottom. The arrangement starts with the lowest-order mode at the top-left corner. Notably, the first nineteen modes appear in direct sequence, whereas the last mode, placed at the bottom right, is specifically the 27th mode. This mode is included over others in the sequence due to its significantly greater impact on wavefront alterations. The disk-shaped discontinuities visible in several modes are the result of the boundary between the primary and tertiary mirrors.}
    \label{fig:m1m3}
\end{figure*}

\section{Related Work} \label{sec:relatedwork}

The literature on the alignment and active optics control of telescopes is as limited as the number of instruments built. The first telescopes to include an AOS were the European Southern Observatory New Technology Telescope \citep{Wilson_1987} and the Very Large Telescopes \citep{Stanghellini_1997}. After these, many telescopes have applied similar approaches, including the Magellan telescopes \citep{Schechter_2003, Palunas_2010} and the Blanco telescope with the Dark Energy Camera (DECam) \citep{Roodman_2014}, among others. Most of these are two-mirror telescopes that only control misalignments of the optical elements and, at most, figure corrections of the primary mirror.

The exploration of degenerate degrees of freedom appears occasionally in the literature. Interestingly, \cite{Sutherland_2015} reporting on the AOS of VISTA, noted that the secondary mirror tilt can be degenerate with primary mirror astigmatism if an insufficient number of sensors are used. More recently, while describing the upcoming improvements to VISTA, \cite{Holzlohner_2022} opted to exclude certain bending modes of their primary mirror to prevent the ill-conditioning of their sensitivity matrix. While they did not address the degeneracy problem per se, it is clear that they are empirically aware of it. In a similar approach, the forthcoming Giant Magellan Telescope (GMT) AOS is planning on excluding certain modes to avoid potential degeneracies \citep{McLeod_2014}. Some of the lowest singular value characteristic modes for GMT for an arbitrary sensitivity matrix were partly described in \cite{Conan_2018}, although no discussion of the effect of noise in inducing degeneracies was included.

In a series of papers, K. Thompson first developed a formalism to express aberration patterns \citep{Thompson_1980, Thompson_2005}. Drawing on that notation, \cite{Schechter_2011} identified the existence of a subspace of benign misalignments in three-mirror anastigmats when only considering coma, astigmatism, and spherical aberration. In this paper, we expand on those findings, although we mostly leverage Zernike polynomials to describe the aberrations.

\begin{figure*}[t!]
    \centering
    \figurenum{2}
    \plotone{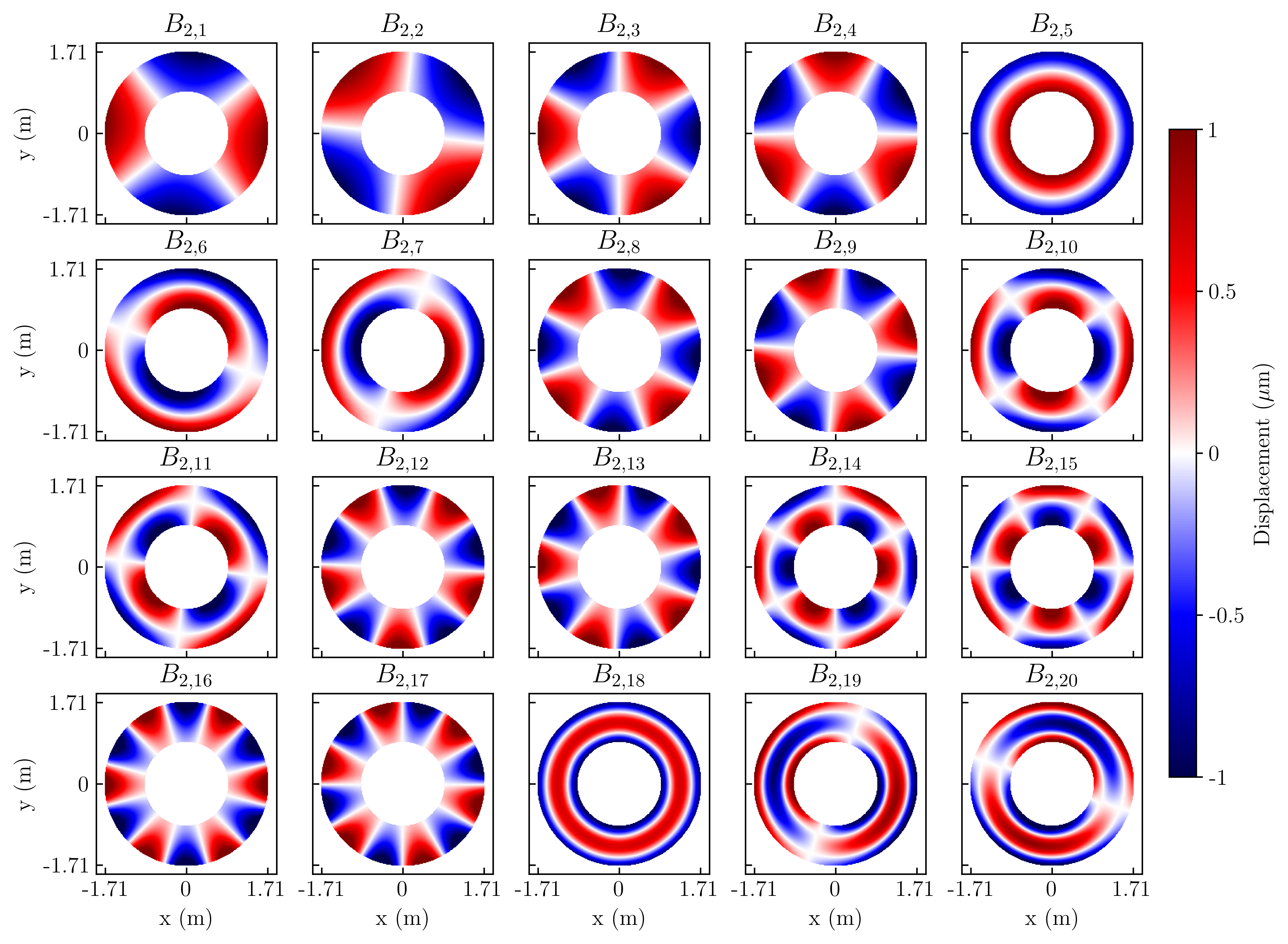}
    \caption{The twenty M2 bending modes used in the Rubin Observatory's AOS, arranged sequentially from left to right and top to bottom. The arrangement starts with the lowest-order mode at the top-left corner. Notably, the first seventeen modes appear in order, whereas the last three modes, placed at the bottom right, are specifically the 25th, 26th and 27th bending modes. These modes are included over others in the sequence due to their significantly greater impact on wavefront alterations.}
    \label{fig:m2}
\end{figure*}

The first reference that discusses aberrations in the Rubin Observatory is in an unpublished M.S. thesis by \cite{Tessieres_2003}. Using Thompson's notation and photon ray-tracing, Tessieres studied the effects of a reduced subset of degrees of freedom in an early model of the Simonyi Survey Telescope. After that, the literature on the Active Optics of the Rubin Observatory is extensive. \cite{Liang_2012} noticed for the first time the ill-conditioning of the sensitivity matrix, without further discussing the issue. Of particular interest is the work by \cite{Angeli_2014} briefly linking the ill-conditioning of the matrix to the presence of noise. While they pointed out the need to remove the five most intuitive degenerate modes, they did not elaborate on the larger subspace of degeneracies introduced by the noise. Similarly, in an internal document, \cite{Neill_2013} observed that atmospheric noise induced excursions in the rigid body motions of the Camera and M2, even when image quality had converged. The document outlined an attempt to mitigate these movements through penalizing them in the control algorithm, yet it did not address the interplay between other degrees of freedom. In more recent papers, the Rubin Observatory-related literature only mentions the degeneracy in passing \citep{Thomas_2023}. Independently, \cite{Yin_2021} studied an alternative approach to the AOS, using Machine Learning to derive the degree of freedom corrections directly from out-of-focus images, but they did not address the effects of the subspace of degeneracies in their implementation.

Some of the techniques that we leverage in this study were inspired by the literature in other fields. Primarily, our methodology makes use of Truncated Singular Value Decomposition (TSVD), for which there is extensive literature on different approaches to determine the singular value cut-off threshold \citep{Falini_2022}. Our proposed truncation criterion, which is based on the compromise between noise error and reconstruction error, is specific to our problem as it needs to be determined from details of how the wavefront errors are measured and characterized. The use of a weighting function to rescale the sensitivity matrix was first introduced by \citep{Kahn_Blisset_1980} in the context of the direct deconvolution of proportional counter X-ray spectra.

Finally, to some extent, the standard notions of model reduction and model balancing from state estimation theory are connected to our work \citep{Pernebo_1982, Moore_1981, Kung_1978}. Our physically-informed rescaling of the sensitivity matrix could be interpreted, in a sense, as a balancing approach that ensures all the linear combinations of degrees of freedom are equally reachable. In the same context, the truncation of the noise-induced degenerate modes can be argued to be similar in spirit to model reduction through balanced truncation \citep{Safonov_1989}. 

\begin{deluxetable}{l r r}[ht]
\tabletypesize{\small}
\tablecaption{\label{table_ranges} Hexapod resolutions and ranges for M2 and Camera.}
\tablecolumns{3}
\tablehead{
\colhead{Component} & \colhead{Resolution} & \colhead{Range} }
\startdata
    M2 Hexapod X/Y axis & 5 $\mu$m & $\pm 6.7$ mm  \\[1pt]
    M2 Hexapod Z axis & 1 $\mu$m & $\pm 5.9$ mm \\[1pt]
    M2 Hexapod Tip / Tilt & $3.3 \cdot 10^{-5}$ deg & $\pm 0.12$ deg \\[1pt]
    M2 Hexapod Rotation in Z & $30 \cdot 10^{-5}$ deg & $\pm 0.05$ deg \\[1pt]
    \tableline  
    Camera Hexapod X/Y axis & 5 $\mu$m & $\pm 7.6$  mm  \\[1pt] 
    Camera Hexapod Z axis & 1 $\mu$m & $\pm 8.7$ mm \\[1pt]
    Camera Hexapod Tip / Tilt & $8.19 \cdot 10^{-5}$ deg & $\pm 0.24$ deg \\[1pt]
    Camera Hexapod Rotation in Z & $60 \cdot 10^{-5}$ deg & $\pm 0.1$ deg \\[5pt]
\enddata
\end{deluxetable}

\begin{deluxetable}{l r r}
\tabletypesize{\small}
\tablecaption{\label{table_mirrors} M1M3 and M2 mirror comparison. M2's higher flexure justifies focusing aberration corrections on M2, especially for noise-prone modes.}
\tablecolumns{3}
\tablehead{
 & \colhead{M1M3} & \colhead{M2} }
\startdata
    Material & Borosilicate glass & ULE  \\[1pt] 
    Thickness & 1 m & 0.1 m \\[1pt] 
    Diameter & 8.4 m & 3.5 m  \\[1pt] 
    Aspect ratio & 8.4 & 35  \\[1pt] 
    Working strength & 100 psi & 1000 psi
\enddata
\end{deluxetable}

\section{Active Optics} \label{sec:active_optics}

The AOS on Rubin is designed to correct the wavefront aberrations induced by gravitational and thermal effects in real-time. This involves sensing aberrations in the wavefront from images and subsequently adjusting the telescope's degrees of freedom, including surface figure modes and misalignments of the optical elements. Typically, an AOS comprises both open and closed loop components, collectively working to minimize the wavefront deviation.

\subsection{Misalignments and Mirror Bending Modes}
To correct the wavefront deviation, the AOS operates with control over a set of degrees of freedom. At the Rubin Observatory, this control extends to fifty degrees of freedom, marking it as one of the first wide-field telescopes with control over a large-dimensional parameter space. This aspect is crucial in understanding the degeneracies we explore herein. These degrees of freedom encompass (1) twenty bending modes ($B_{1, i}$) for the primary/tertiary mirror (M1M3), as shown in Figure \ref{fig:m1m3}, (2) twenty additional bending modes ($B_{2, i}$) for the secondary mirror (M2), (3) the rigid body motions of M2, including decentering, piston, tip, and tilt, and (4) analogous rigid body motions for the Camera.

\begin{deluxetable}{l r r}
\tabletypesize{\small}
\tablecaption{\label{table_bending} Available mirror bending mode ranges derived by uniformly distributing the actuator force range across all bending modes to ensure equal force allocation per mode. The range per mode is calculated from this allocated force using the influence matrix that relates actuator forces (N) to bending modes ($\mu$m in surface deviation).}
\tablecolumns{3}
\tablehead{
\colhead{Bending mode} & \colhead{M1M3 ($j = 1$)} & \colhead{M2 ($j = 2$)} }
\startdata
    $B_{j,1}$ & $\pm$ 0.454 $\mu$m & $\pm$ 4.287 $\mu$m  \\[1pt] 
    $B_{j,2}$ & $\pm$ 0.452 $\mu$m & $\pm$ 4.306 $\mu$m  \\[1pt] 
    $B_{j,3}$ & $\pm$ 0.087 $\mu$m & $\pm$ 0.609 $\mu$m  \\[1pt] 
    $B_{j,4}$ & $\pm$ 0.066 $\mu$m & $\pm$ 0.557 $\mu$m  \\[1pt] 
    $B_{j,5}$ & $\pm$ 0.066 $\mu$m & $\pm$ 0.331 $\mu$m  \\[1pt]
    $B_{j,6}$ & $\pm$ 0.023 $\mu$m & $\pm$ 0.136 $\mu$m  \\[1pt] 
    $B_{j,7}$ & $\pm$ 0.021 $\mu$m & $\pm$ 0.138 $\mu$m  \\[1pt] 
    $B_{j,8}$ & $\pm$ 0.022 $\mu$m & $\pm$ 0.160 $\mu$m  \\[1pt] 
    $B_{j,9}$ & $\pm$ 0.019 $\mu$m & $\pm$ 0.159 $\mu$m  \\[1pt] 
    $B_{j,10}$ & $\pm$ 0.013 $\mu$m & $\pm$ 0.076 $\mu$m  \\[1pt] 
    $B_{j,11}$ & $\pm$ 0.013 $\mu$m & $\pm$ 0.075 $\mu$m  \\[1pt] 
    $B_{j,12}$ & $\pm$ 0.009 $\mu$m & $\pm$ 0.064 $\mu$m  \\[1pt] 
    $B_{j,13}$ & $\pm$ 0.009 $\mu$m & $\pm$ 0.065 $\mu$m  \\[1pt] 
    $B_{j,14}$ & $\pm$ 0.009 $\mu$m & $\pm$ 0.039 $\mu$m  \\[1pt] 
    $B_{j,15}$ & $\pm$ 0.004 $\mu$m & $\pm$ 0.033 $\mu$m  \\[1pt] 
    $B_{j,16}$ & $\pm$ 0.004 $\mu$m & $\pm$ 0.030 $\mu$m  \\[1pt] 
    $B_{j,17}$ & $\pm$ 0.006 $\mu$m & $\pm$ 0.032 $\mu$m  \\[1pt] 
    $B_{j,18}$ & $\pm$ 0.006 $\mu$m & $\pm$ 0.011 $\mu$m  \\[1pt] 
    $B_{j,19}$ & $\pm$ 0.005 $\mu$m & $\pm$ 0.008 $\mu$m  \\[1pt] 
    $B_{j,20}$ & $\pm$ 0.002 $\mu$m & $\pm$ 0.007 $\mu$m 
\enddata
\end{deluxetable}

Control over the bending modes of M1M3 is facilitated by 156 force actuators distributed across the mirror cell, with an influence matrix encoding the correspondence between each bending mode and the applied forces. Similarly, M2 bending modes are controlled by 72 axial actuators and 4 tangential actuators, while misalignments for M2 and the Camera are controlled through two independent hexapod mechanisms.

\begin{figure*}[t!]
    \centering
    \figurenum{3}
    \epsscale{1.15}
    \plotone{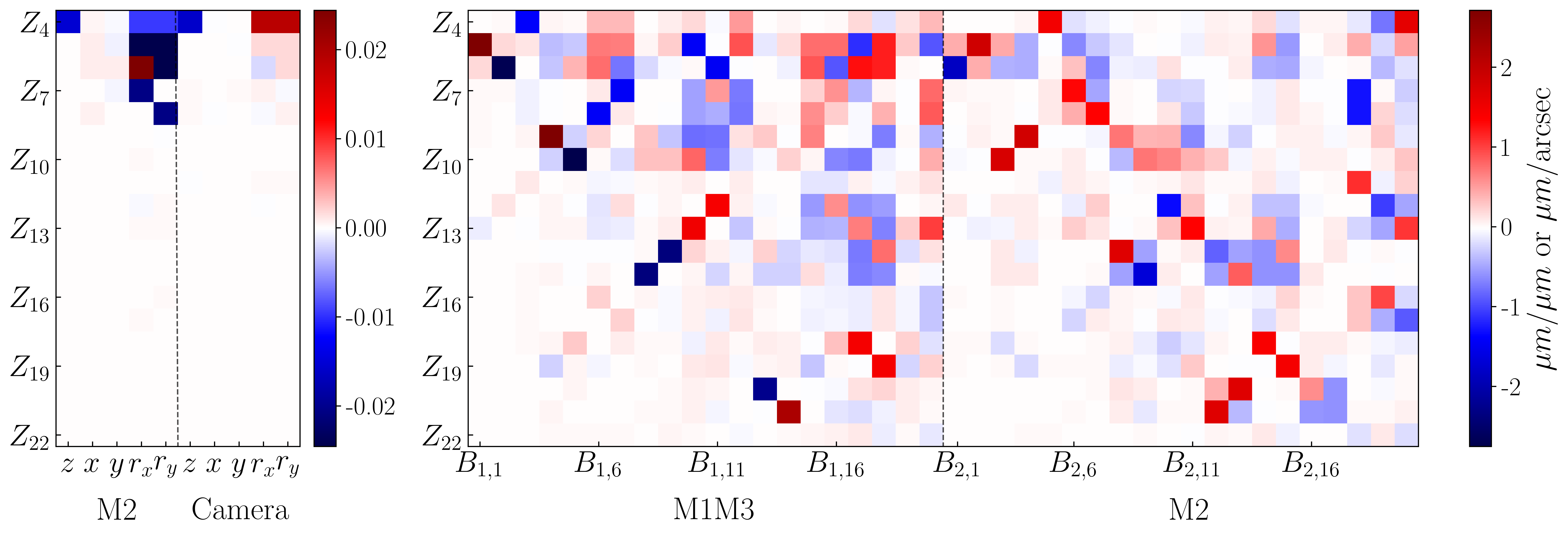}
    \caption{Sensitivity matrix in the corner R00 of the Rubin Observatory focal plane. Due to the scaling difference between the effect of hexapod motions and bending modes, they have been split in two.}
    \label{fig:sensitivity_matrix}
\end{figure*}

The available operational ranges for AOS corrections for the different degrees of freedom are detailed in Table \ref{table_ranges} and Table \ref{table_bending}. The ranges available for AOS corrections on bending modes are derived from the single actuator force range which is $\pm 22.5 \mathrm{N}$ for M2 actuators and $\pm 67 \mathrm{N}$ for M1M3 actuators. This range is distributed uniformly accross the different bending modes and the final range is obtained from the forces required by each bending mode. Notably, as seen in Table \ref{table_bending}, M2 bending mode ranges allow for a far larger movement than M1M3. Additionally, Table \ref{table_mirrors} provides an overview of the physical attributes of the primary/tertiary and secondary mirrors, highlighting the enhanced flexure exhibited by M2 in comparison to M1M3. These ranges serve as a relevant reference for the subsequent scaling approach introduced in our study.

\subsection{Open loop component}
The open loop component utilises a Look-Up Table (LUT) containing the degrees of freedom corrections that can be predicted from aberrations caused by temperature, azimuth, elevation, and camera rotation. At the Rubin Observatory, this LUT is encoded using independent fifth-order polynomials, each accounting for a different variable. However, certain aberrations induced by these variables remain unpredictable and change over time. Among these residual aberrations, those that change slowly in time, in particular the ones due to thermal gradients and wind, can be addressed by the closed-loop component of the AOS, which utilizes wavefront error estimates to compute additional corrections.

\subsection{Closed loop component}

The closed-loop component adds a final layer of control to address slowly changing aberrations not corrected by the Look-Up Table (LUT). It accomplishes this by utilizing defocused images to estimate the wavefront at the pupil, which in turn informs the optical state of the telescope, including residual misalignments and optical surface errors. From this optical state estimate, the degree of freedom corrections are derived.

At the Rubin Observatory, wavefront sensing occurs at the four corners of the focal plane using a $\pm 1.5 \ \mathrm{mm}$ intra and extra-focal CCD pair, producing out-of-focus ``donuts''. These donuts are then used to estimate the wavefront in Zernike coefficients through methods such as solving the Transport of Intensity Equation (TIE) \citep{Xin_2015, SITCOMTN-111}, forward modeling techniques like Danish \citep{Janish_2012}, or Neural Networks trained on simulated images \citep{DThomas_2020, Crenshaw_2024}.

Subsequently, the closed-loop system derives the optical state from the Zernike estimate through the inversion of the sensitivity matrix. This matrix quantifies how the wavefront, expressed in Zernike coefficients, changes with the movement of each degree of freedom by one unit (either, one micron for translations or one arcsecond for rotations). In this definition, there is an implicit linearization assumption that only holds for small movements, which is generally valid as long as the LUT primarily handles the major aberration corrections. Additionally, the sensitivity matrix is defined at each focal plane position due to the dependence of the wavefront on pupil and focal plane coordinates. In this study, we focus on the sensitivity matrix at the four corners of the focal plane. An example of the sensitivity matrix with effects of each degree of freedom in one of the Camera corners is illustrated in Figure \ref{fig:sensitivity_matrix}.

The choice of Zernike polynomials is primarily for convenience and alternative bases could be chosen. Regardless of the chosen basis, there is not a one-to-one mapping between the optical state of the telescope and the wavefront estimate. Therefore, it is important to directly asses the accuracy of optical state estimation in terms of degrees of freedom.

Once the estimate of the wavefront deviation is obtained, the sensitivity matrix is ``inverted'' to determine the estimated corrections to the degrees of freedom. In the prior Rubin AOS software system, the final applied corrections were subsequently derived through an OIC controller \citep{Neill2014, SITCOMTN-129}. The formulation and details of this process are discussed in the following section, along with the exploration of degeneracies in both state estimation and correction calculations.

\section{Formulation} \label{sec:formulation}
To estimate the optical state of the system, we utilize the wavefront deviation, computed by subtracting the reference wavefront \textemdash the inherent optical aberrations that cannot be corrected \textemdash from the wavefront error estimates and represented through Zernike coefficients. We model our system as a discrete-time linear dynamic system through the application of a sensitivity matrix, outlined by the equations:
\begin{equation}
\begin{aligned}
    {x}_{k + 1} = {x}_{k} + {\mu}_{k} + {\delta}_{k} \\ 
    {y}_{k} = A {x}_{k}
\end{aligned}
\end{equation}

Here, $y_k \in \mathbb{R}^n$ denotes the wavefront deviation in Zernike coefficients (measured in micrometers), $x_k \in \mathbb{R}^m$ symbolizes the system's optical state in the degrees of freedom basis (measured in micrometers for bending modes and displacements, and arcseconds for tips and tilts), $\mu_k$ represents the control correction vector in the same basis, and $\delta_k$ accounts for external disturbances between exposures. The matrix $A \in \mathbb{R}^{(n,m)}$ is the sensitivity matrix with appropriate units. Specifically for the Rubin Observatory, the wavefront is estimated in four corners of the focal plane, with each estimate consisting of 19 annular Zernike coefficients\textemdash from Zernike 4th (defocus) to the 22nd (secondary spherical), using Noll's index notation\textemdash, yielding $n = 76$. The optical state is then determined by estimating $m = 50$ degrees of freedom from the 76 Zernike coefficients. For a summary of the notation used in this paper, see Table \ref{notation}.

In control theory terms, this setup is deemed controllable and observable. Nevertheless, the introduction of noise in the wavefront estimates compromises the sensitivity matrix's rank, impacting full observability. To derive the necessary corrections for the telescope, we first estimate $x_k$ based on the wavefront deviation $y_k$, and then apply a standard controller to adjust the telescope's optics. Prior to our work, Rubin used an Optimal Integral Controller (OIC) derived from solving an optimal control problem, which cost function balanced image quality and incremental movement of degrees of freedom. Despite achieving optimal image quality, the OIC method encounters difficulties due to inherent system degeneracies, leading to excursions within the noise-induced subspace of degeneracies. The focus of our study is therefore to develop a solution that minimizes these deviations, while maintaining the image quality.

\begin{deluxetable}{l l}
\tabletypesize{\small}
\tablecaption{\label{notation} Summary of key notation used in this paper.}
\tablecolumns{2}
\tablehead{Parameter & Description }
\startdata
   $x$ & Optical state \\[1pt]
   $\hat{x}$ & Optical state estimate \\[1pt]
    $y$ & Wavefront deviation \\[1pt] 
    $A$ & Sensitivity matrix  \\[1pt] 
    \tableline  
    $\mu_k$ & Control corrections \\[1pt] 
    $K_p$, $K_i$, $K_d$ & Proportional, integral and derivative gains \\[1pt] 
    $w$ & Noise contribution \\[1pt] 
    $S_i$ & Spectral power contribution \\[1pt]  
    \tableline  
    $\Sigma$ ($\sigma_i$) & Singular value matrix  \\[1pt] 
    $V$ ($v_i$) & Right singular vectors matrix  \\
    [1pt] 
    $U$ ($u_i$) & Left singular vectors matrix  \\[1pt] 
    \tableline  
    $\Omega$ & Re-scaling matrix  \\[1pt]
    $\tilde{\Sigma}$ ($\tilde{\sigma}_i$) & Re-scaled singular value matrix  \\[1pt] 
    $\tilde{V}$ ($\tilde{v}_i$) & Re-scaled right singular vectors matrix  \\[1pt] 
    $\tilde{U}$ ($\tilde{u}_i$) & Re-scaled left singular vectors matrix  \\[1pt] 
\enddata
\end{deluxetable}

\subsection{Optical state estimation}
The focal point of our analysis is the estimation of the optical state $x$ from the wavefront deviation measurement $y$, a process we will refer to as optical state estimation. For the sake of simplicity we henceforth omit the time index $k$. Given the sensitivity matrix is not square, we use the Moore-Penrose pseudo-inverse ($A^{\plus}$) or, equivalently, employ a linear square minimization approach. This can be described using SVD matrices as follows:

\begin{equation}
    \hat{x} = A^{\plus} y =  V \Sigma^{-1} U^{\tr} y = \sum_{i=0}^m \frac{\langle y, u_i \rangle}{\sigma_i}v_i
\end{equation}

\begin{figure*}[ht!]
    \centering
    \figurenum{4}
    \epsscale{1.1}
    \plotone{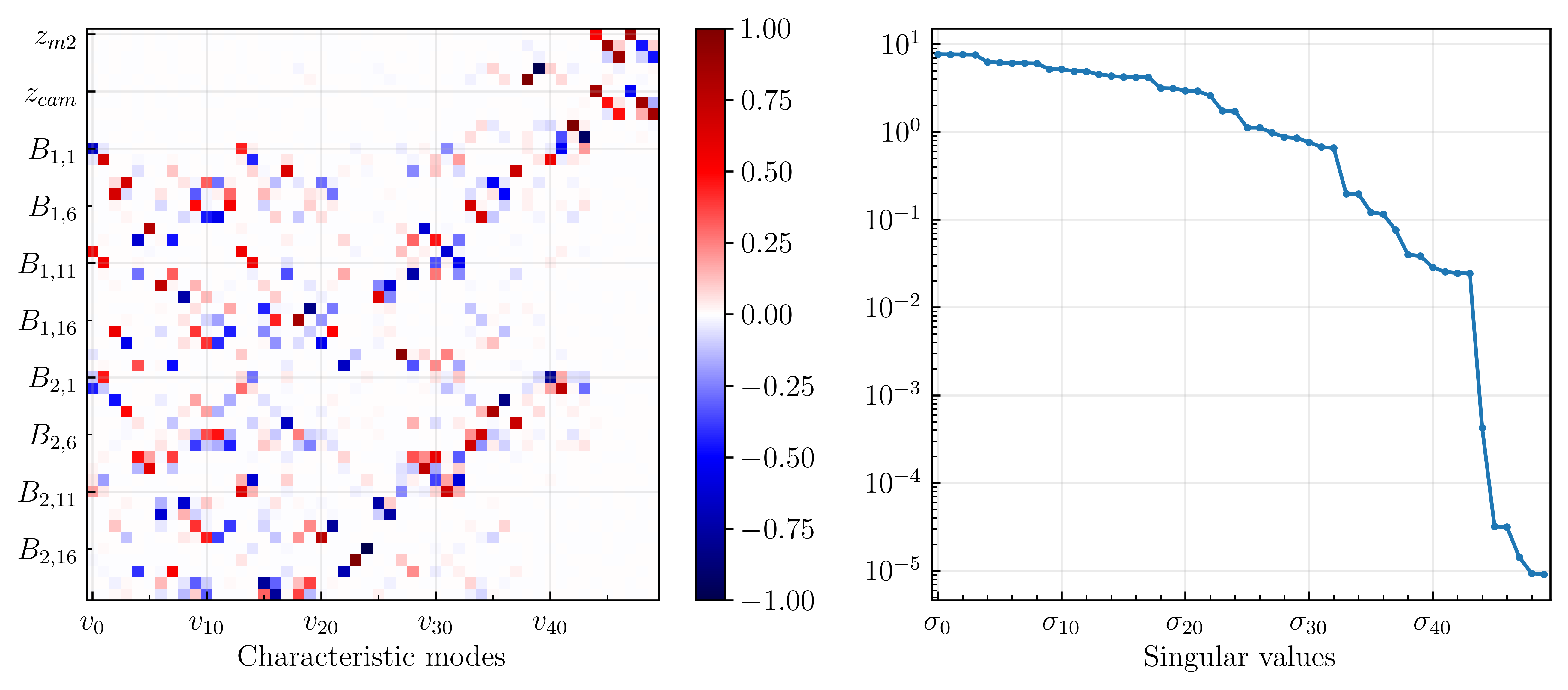}
    \caption{(Left) Right singular vectors of the original sensitivity matrix at the four corners, organized from largest to smallest singular value. The y-axis represents the 5 rigid body motions of M2, the 5 motions of the camera, bending modes of M1M3, and bending modes of M2. The colormap reflects the respective units of each degree of freedom, measured in $\mu$m or arcsec. (Right) Singular values associated with each mode. A distinct set of singular values is significantly lower, corresponding exclusively to the rigid body motions.}
    \label{fig:unweighted_svd}
\end{figure*}

Here, under the condition that $m<n$, $U \in \mathbb{R}^{(n,m)}$ is a semi-unitary matrix representing the left singular vectors ($u_i$) of the sensitivity matrix; $\Sigma \in \mathbb{R}^{(m,m)}$ is the diagonal matrix containing the singular values $\sigma_i$, arranged in descending order; and $V \in \mathbb{R}^{(m,m)}$ is a unitary matrix comprising the right singular vectors $v_i$ in the basis of degrees of freedom. The vectors $v_i$ denote the telescope's \textit{characteristic modes}, essentially linear combinations of the degrees of freedom. An illustration of $V$ and $\Sigma$ is provided in Figure \ref{fig:unweighted_svd}.

This method of inversion is ideal when matrix $A$ possesses full rank. In a noise-free environment, this enables precise estimates of the corrections. However, noise impacts this inversion process, effectively reducing the rank of $A$. An early indication of this effect can be gleaned by evaluating the condition number of matrix $A$, defined as
\begin{equation}
    \kappa (A) = \frac{\sigma_{max}}{\sigma_{min}}.
\end{equation}
For the Rubin Observatory, the condition number is approximately $8.4 \cdot 10^5$. This high value indicates a significant sensitivity to noise: even small errors in the wavefront measurement, on the order of 1 micron, can lead to errors in the optical state of approximately $8.4 \cdot 10^5$ microns (or equivalent in arcseconds). This underscores the need to effectively mitigate noise, which is further explored in the following section.

\subsection{Noise effects}
Signal-to-noise ratio (SNR), atmospheric turbulence, and inaccuracies in wavefront estimation lead our wavefront error measurements to deviate from their true values. To account for this, we can modify the output equation of our system to include a noise component on the wavefront deviation, expressed as:
\begin{equation}
    y = Ax + w
\end{equation}

Here, $w \in \mathbb{R}^n$ represents the noise in the wavefront error measurements, with units in micrometers of Zernike coefficients. It is assumed to be well-behaved in accordance with the Central Limit Theorem. Incorporating this noise, our estimation becomes: 
\begin{equation}
    \hat{x}_{noisy} =  A^{\plus} y +  A^{\plus} w
\end{equation}

The noise-induced error, when compared to an ideal estimation, is then constrained by:
\begin{equation}
    ||\hat{x}_{noisy} - \hat{x}||^2_2 \leq \max_{\sigma_i} \frac{1}{\sigma_i^2} ||w||_2^2
\end{equation}
This error is limited by the smallest singular value, indicating that even minor noise can significantly increase the error, especially when small singular values exist \citep{Moon_2000}. Figure \ref{fig:unweighted_svd}, Rubin's sensitivity matrix, exhibits five modes with notably low singular values and is especially prone to such error amplification, even at low noise levels. To mitigate this, discarding singular values most affected by noise, a method known as Truncated SVD (TSVD), becomes essential. This process reduces the operational basis to a non-degenerate set of characteristic modes.

Determining the truncation threshold requires establishing the extent of the noise-induced subspace of characteristic modes which have a negligible impact on the wavefront and are inherently degenerate. In this study, we employ simulated images and a power spectrum approach to discern the onset of these noise-induced degeneracies. This analysis reveals the noise dominance threshold in our Zernike estimates. For that purpose, it is useful to define the power spectrum of the wavefront deviation estimates, utilizing the expanded SVD basis, where $U \in \mathbb{R}^{(n,n)}$,
\begin{equation}\label{power_eq}
     S_i =  \frac{\langle y, u_i \rangle ^2}{\sum_i \langle y, u_i \rangle^2}
\end{equation}
Here, $S_i$ quantifies the contribution of each mode $u_i$ to the total wavefront deviation.

This methodology enables the identification of noise thresholds beyond which Zernike estimates are predominantly noise-influenced, guiding the truncation of singular values. However, truncation introduces an approximation error corresponding to the optical state's projection onto the incomplete basis of truncated singular vectors. To better understand the impact of this approximation, we introduce an effective response matrix $V V^{\tr}$, where 
$V$ represents the SVD matrix of singular vectors after truncation. This matrix quantifies the approximation error by projecting onto the truncated singular vectors basis and is applied as follows:
\begin{equation}
    \hat{x} =  A^{\plus} A x = V V^{\tr} x = \sum_{i = 0}^{K} \langle x, v_i \rangle v_i
\end{equation}
Here, K denotes the number of non-truncated singular vectors, with $K < m$, indicating a reduction in dimensionality.

The effective response matrix is crucial for comparing the actual state of the telescope with the estimates produced by the method. An example of this matrix after truncation is illustrated in Figure \ref{fig:vvt}. Truncating the smallest singular values, a step necessitated by noisy conditions, results in the creation of off-diagonal elements, which implies that perturbations in particular degrees of freedom can be mapped into others. This process also reveals the degenerate characteristic modes within the matrix, providing insights into the system's behavior under noise influence.

\subsection{Re-scaled Sensitivity Matrix}
Analyzing the singular values in Figure \ref{fig:unweighted_svd} uncovers significantly small singular values corresponding to the decenterings of M2 and the camera, as well as the opposite pistoning movements of both. However, as we discuss below, the impact of noise extends beyond these five lowest singular values. Considering that the contributions from all rigid body motions appear at the lower end of the singular value spectrum, simply truncating these values would result in the complete exclusion of the rigid body adjustments from the solution. Such an approach is not advisable, since rigid body adjustments are actually the easiest to make in the system, given the large operational ranges available to the actuators that control them. The problem stems from the arbitrary nature of units, ranges, and scales used in the sensitivity matrix.

To reconcile these discrepancies among different degrees of freedom, thereby enabling the analysis of couplings between misalignments and bending modes, we introduce a rescaling of the sensitivity matrix. This approach was first introduced by \citep{Kahn_Blisset_1980} in another context. The rescaling affects the characteristic mode basis and hence the subspace of degeneracies. With a rescaling matrix $\Omega$, which we will take to be diagonal, the output equation of our system can be written as,
\begin{equation}
    y = A \Omega \  (\Omega^{-1} x) = \tilde{A} (\Omega^{-1} x)
\end{equation}

Through this normalization, the effective response matrix is transformed to:
\begin{equation}
\begin{split}
    \hat{x} 
    & =  \Omega \tilde{A}^{\plus} \tilde{A} \Omega^{-1} x = \Omega \tilde{V} \tilde{V}^{\tr}\Omega^{-1}  x \\ 
    &= \sum_{i = 0}^{K} \langle \Omega^{-1}  x, \tilde{v}_i \rangle \Omega \tilde{v}_i
\end{split}
\end{equation}
Here, $\tilde{A}$ denotes the rescaled matrix, and $\tilde{V}$ and $\tilde{U}$ the resulting rescaled basis, with their corresponding rescaled vectors expressed as $\tilde{v}_i$ and $\tilde{u}_i$. 

In the course of this study, we explored a variety of rescaling options and found the most effective approach to be weighting by the effective wavefront error range attributable to each degree of freedom, given the system's constraints. This involves using the operational range of the degrees of freedom $\Delta x_i$, as presented in Table \ref{table_ranges} and Table \ref{table_bending}, and the impact of each degree of freedom on the Full-Width Half Maximum (FWHM) of the images $\Delta \psi_i$. Thus, we define $\Omega$ as the following diagonal matrix:
\begin{equation}\label{weight_equation}
    \Omega_{i,i} = \Delta x_i \Delta \psi_i
\end{equation}
Here, $\Omega_{i,i}$ refers to the diagonal elements, with one such term corresponding to each of the $m$ degrees of freedom. The term $\Delta x_i$ denotes the operational range of the $i$-th degree of freedom, measured either in $\mu$m or arcsec, depending on the context. The factor $\Delta \psi_i$ quantifies the effect on the FWHM, averaged over the field and measured in arcseconds, of altering the $i$-th degree of freedom by one unit of its operational range. Consequently, $\Delta \psi_i$ has units of arcsec/$\mu$m or arcsec/arcsec, reflecting the ratio of the change in FWHM to the change in the operational range. The resulting $\Omega_{i,i}$ is expressed in units of arcsec, and is proportional to the full range of FWHM variation for each degree of freedom.

This rescaling strategy not only facilitates a more nuanced understanding of the effect of the various degrees of freedom, but it also refines our approach to mitigating the effects of noise and degeneracy within the system.

\subsection{Controller}
After estimating the optical state, the remaining step involves calculating corrections to be implemented. This can be approached through two standard methods in the field. The OIC controller previously planned for Rubin was designed to minimize the cost function:
\begin{equation}
    J = \sum_{i\in\mathcal{O}} w_i (x_{k+1}^{\tr} Q x_{k+1})_i + \rho^2 u^{\tr} H u,
\end{equation} 
where $Q$ represents the image quality matrix, assessing the impact of the optical state $x$ on the Normalized Point-Source Sensitivity (PSSN), $H$ corresponds to a penalty matrix, $\rho$ is the penalty gain, and the sum is averaged over a set of Gauss-Legendre points defined on the focal plane.

In the new approach we present here, the corrections are calculated through a simple PID control loop that uses the estimated optical state as the error value, a method that has been used by multiple previous telescopes \citep{Roodman_2014}. This approach adjusts the telescope to minimize the derived optical state. The control actions in this model are defined as:
\begin{equation}
\begin{split}
    \mu_k & = K_p (x_k - x_{k-1}) + K_iT_s x_k \\ & + K_d \frac{x_k - 2x_{k-1} + x_{k-2}}{T_s}
\end{split}
\end{equation}
Here, $K_p$, $K_i$, $K_d$, are the proportional, integral ad derivative gains, respectively. In our implementation, we propose the use of the reduced basis optical state estimate to guide the PID loop corrections, therefore excluding those characteristic modes rendered degenerate due to noise.

\section{Simulations} \label{sec:simulations}
In this section, we detail the simulation tools employed to conduct our study for the Rubin Observatory, focusing on the impact of noise across various optical states. Our simulation framework integrates advanced photon-ray tracing and atmospheric modeling techniques, utilizing the Batoid, GalSim, and ImSim python packages for a comprehensive analysis.

The optical model of the Rubin Observatory telescope is simulated using Batoid, as outlined by \cite{Meyers_2019}, providing a high-fidelity representation of the telescope's optics. For atmospheric effects and closed-loop simulations, we employ GalSim \citep{Galsim_2016} for atmosphere modeling and ImSim \citep{imSim} for dynamic closed-loop simulation scenarios.

Our analysis investigates the influence of noise on 100 distinct optical states. These states mimic realistic degree of freedom configurations by uniformly distributing the effects of the total optical aberration across the telescope's degrees of freedom, in line with the methodology used by \cite{Crenshaw_2024}. The total optical aberration is drawn from a truncated Gaussian with a mean of 0.8 arcsec of FWHM. Each state is subjected to a unique atmospheric condition, modeled using GalSim's implementation of the von Karman turbulence model \citep{Karman_1948} alongside the Ellerbroek model \citep{Ellerbroek_2002}. This approach simulates the atmospheric structure through a series of frozen phase screens, each representing the turbulence between two altitudes and moved by specific wind velocity vectors. For each optical state, we simulate a pair of non-vignetted, 14th magnitude intra- and extra-focal donut images at the center of each of the wavefront sensors located at the four corners of the focal plane. Using the TIE baseline method employed by the Rubin Observatory, we estimate the wavefront error. Subsequently, we compute the wavefront deviation by subtracting this estimated wavefront from the reference wavefront, which represents the inherent optical aberrations that cannot be corrected. The simplifications made to simulate the donuts are primarily to minimize the computational time required for simulating images with multiple sources. In reality, the imperfections in actual donuts could lead to worse wavefront estimates than those simulated.

The sensitivity matrix, crucial for interpreting the telescope's response to various aberrations, is derived from a double Zernike model generated with Batoid which combines the pupil and field dependency in one expansion \citep{Kwee_1993}. This model's accuracy is within $1\%$ of the mirror acceptance testing results, aligning with expectations for the real system's performance. Our simulations are primarily conducted in the $r$ band at a wavelength of 622 nm, although similar methodologies can be applied across other filters for broader analysis.

For closed-loop simulations, we utilize the ImSim package to incorporate adjustments in the telescope's elevation (and resulting gravitational flexure) between exposures and to factor in inherent LUT errors estimated at $1\%$. These simulations leverage the optical feedback controller python package from the Rubin Observatory, allowing for a detailed exploration of the system's dynamic response to corrections. The different consecutive elevations used in these dynamic simulations are taken from the LSST baseline survey.

\section{Results} \label{sec:results}

In this section, we present our findings on the subspace of degeneracies for the Rubin Observatory and assess the impact of employing an improved optical state estimate on the closed-loop performance for a simulated run of the LSST. Initially, we examine and compare the characteristic modes derived from the original unscaled sensitivity matrix against those from our rescaled sensitivity matrix, including a discussion on the nature of each characteristic mode.

\begin{figure*}[t!]
    \centering
    \figurenum{5}
    \epsscale{1.1}
    \plotone{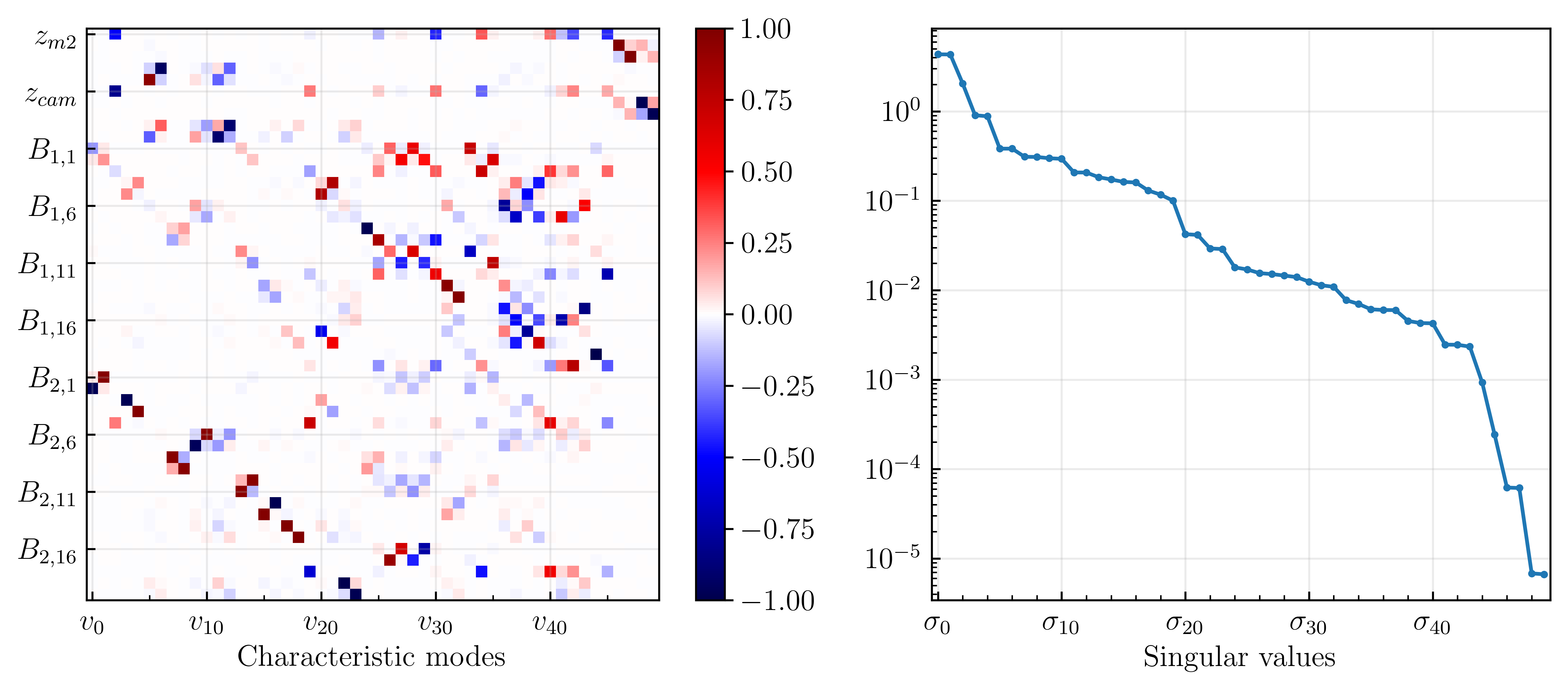}
    \caption{(Left) Right singular vectors of the rescaled sensitivity matrix at the four corners, organized from largest to smallest singular value. The y-axis represents the 5 rigid body motions of M2, the 5 motions of the camera, bending modes of M1M3, and bending modes of M2. The colormap reflects the respective units of each degree of freedom, measured in $\mu$m or arcsec. (Right) Singular values associated with each mode.}
    \label{fig:weighted_svd}
\end{figure*}

Following this comparison, we present the power spectrum plots that guide the determination of the truncation thresholds for the rescaled characteristic mode basis. We then illustrate how applying these thresholds modifies the effective response matrix in both scenarios.

Concluding our analysis, we evaluate the performance of the previous OIC in comparison to a PID control loop, which has been refined based on insights from our noise study and the application of TSVD. This comprehensive examination aims to illuminate the benefits of an improved optical state estimation for enhancing the overall system performance of the Rubin Observatory.

\subsection{Characteristic modes}

In our analysis, we first examine the differences between characteristic modes derived from the unscaled and scaled sensitivity matrices. Figure \ref{fig:unweighted_svd} displays the characteristic modes and singular values for the unscaled sensitivity matrix, whereas Figure \ref{fig:weighted_svd} presents the modes post-scaling, as dictated by the weights specified in Equation \ref{weight_equation}. 

Significant differences are evident between the two resulting bases. Notably, the unscaled matrix demonstrates a clear demarcation between bending modes and rigid body motions, as illustrated in Figure \ref{fig:unweighted_svd} by its nearly block-diagonal structure. Meanwhile, the scaled sensitivity matrix shows the interplay between both sets of degrees of freedom. This split in the unscaled basis is attributable to the distinct effects that a unit displacement has on the two sets of degrees of freedom. Indeed, a 1 $\mu$m shift in the rigid body motions exerts a minimal impact on the wavefront, in contrast to the significant influence of a 1 $\mu$m movement in the bending modes. This results in rendering all the rigid body motions irrelevant in the unscaled case, as evidenced by the upper-diagonal block characterized by the lowest singular values within the unscaled matrix. Scaling adjusts for this disparity by weighting modes according to the available wavefront's FWHM range that each degree of freedom can afford given the operational ranges, thereby ranking the SVD by their influence. Tables \ref{table_bending} and \ref{table_ranges} elucidate this further: hexapod movements permit extensive correctional range, unlike the more restricted bending modes. Although the operational ranges shown in these tables inherently encapsulate available motion, direct weighting by them would overestimate the impact of modes that, despite substantial range, minimally affect the wavefront. Thus, scaling by FWHM available range incorporates a more nuanced understanding of the degree of freedom significance.

Further inspection of the characteristic modes of both matrices reveals similarities among the five modes with the lowest singular values. The least impactful on the wavefront, in both instances, are the opposing decenterings of M2 and the Camera. While the unscaled matrix singles out the pistoning of the Camera and M2 in opposite directions in $v_{48}$, scaling introduces a mode $\tilde{v}_{45}$ that couples this action with $B_{1,3}$, $B_{1,12}$, and $B_{1,20}$ (shown in Figure \ref{fig:m1m3}), which collectively mimic M1M3 pistoning. It also includes to a lesser degree bending modes $B_{2,5}$ and $B_{2,18}$, analogous modes for M2 as shown in Figure \ref{fig:m2}. Finally, the other two characteristic modes with the least effect on the wavefront correspond to concurrent decenterings in $x$ and $y$ axis for both the Camera and M2, which, due to their minimal impact on the wavefront within their operational range, are deemed irrelevant even in the scaled case.

Furthermore, the unscaled matrix identifies the concurrent pistoning of M2 and the Camera in the same direction, along with tips and tilts, as minimally impactful, a finding not mirrored in the scaled matrix. Specifically, the scaled analysis introduces concurrent pistoning of the camera and M2 coupled with $B_{2,5}$ and $B_{1,3}$ as the third-most significant characteristic mode. Again, this difference in the scaled case is due to the introduction of the coupling between pistonings and the corresponding bending modes across both mirrors. For rotations, their prominence as relevant characteristic modes on the scaled matrix arises from addressing and removing arbitrary unit discrepancies within the sensitivity matrix.

Another inherent system degeneracy involves the interplay between $B_{1,2}$ and $B_{2,1}$, and between $B_{1,1}$ and $B_{2,2}$. When these are actuated in opposing directions, their effects neutralize, whereas concurrent movements result in additive effects. These two modes are apparent in the unscaled version on $v_0$ and $v_{41}$, and $v_1$ and $v_{40}$. After scaling, however, the characteristic modes with high singular values tend to emphasize M2 bending modes instead of M1M3 ones, thus rendering $B_{1,1}$ and $B_{1,2}$ comparatively less significant in favor of $B_{2,1}$ and $B_{2,2}$. This effect is also manifest at the spectrum's lower end: while the unscaled matrix identifies the tenth and ninth most degenerate characteristic modes ($v_{40}$ and $v_{41}$) as involving the opposite movements of $B_{1,2}$ and $B_{2,1}$, and $B_{1,1}$ and $B_{2,2}$, scaling adjusts their significance to $\tilde{v}_{33}$ and $\tilde{v}_{35}$, incorporating a pronounced contribution from M1M3 modes.

Lastly, mode $\tilde{v}_{43}$ in the rescaled basis, combining $B_{1,6}$, $B_{1,15}$, and to a lesser extent $B_{1,13}$, $B_{1,17}$, and $B_{2,7}$, also warrants mention. The bending mode Figures \ref{fig:m1m3} and \ref{fig:m2} suggest why these combinations minimally affect or neutralize each other's impact on the wavefront. A similar rationale, grounded in optical intuition, can be applied with varying degrees of ease to the remainder of the characteristic modes to discern the interplay among different degrees of freedom.

Examining these results, it becomes clear the characteristic modes do not necessarily represent the actual state of the telescope's optics but instead indicate the ``easiest'' way, within the system's limitations, of producing the measured wavefront errors. This will not guarantee the precision of the estimated optical state; however, the contributions from the estimated state will provide a comparable wavefront and are the best estimate and strategy for correction we can derive. For example, in the presence of noise, it's impossible to determine whether the optical aberration caused by $B_{1,1}$ originates from M1M3 or its counterpart $B_{2,2}$ on M2. This ambiguity underscores the value of characteristic modes as a crucial framework for estimating the optical state and implementing system corrections. Indeed, the rescaled characteristic mode basis selects M2 as the prevalent mode, reflecting its greater corrective capacity within the system. 

Finally, it is worth mentioning the changes observed in the singular value distribution. For the rescaled matrix we still have significantly lower singular values, but the decay is steeper, effectively resembling a power law, an effect arising from the rescaling of the different modes.

\begin{figure*}[h]
    \centering
    \figurenum{6}
    \epsscale{1.1}
    \plotone{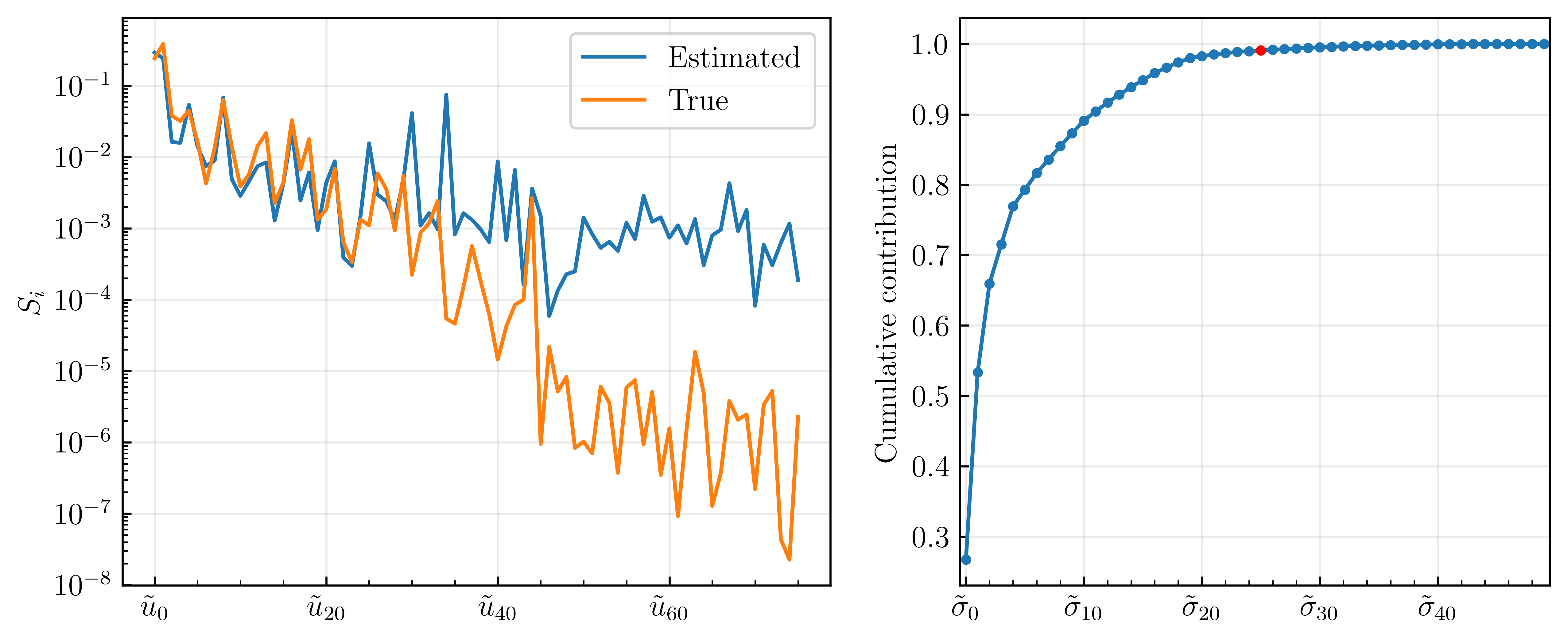}
    \caption{(Left) This panel illustrates a comparison between the power spectrum $S_i$ of the true wavefront deviation and the estimated deviation. The line plots represent the averages across 100 distinct optical states used to perturb the telescope, simulating for each state corner out-of-focus images under different atmospheric instances, each characterized by an average seeing of 0.8 arcsec. The x-axis delineates the various $\tilde{u}_i$ modes within the wavefront deviation basis, encompassing 76 modes derived from the extended $\tilde{U}$ basis via SVD. (Right) Here, we display the cumulative sum of singular values. The red dot marks the juncture where the cumulative contribution attains 99\%, highlighting the significant portion of the total contribution captured by a subset of the modes.}
    \label{fig:power_spectrum}
\end{figure*}

\begin{figure*}[h]
    \centering
    \figurenum{7}
    \epsscale{1.15}
    \plotone{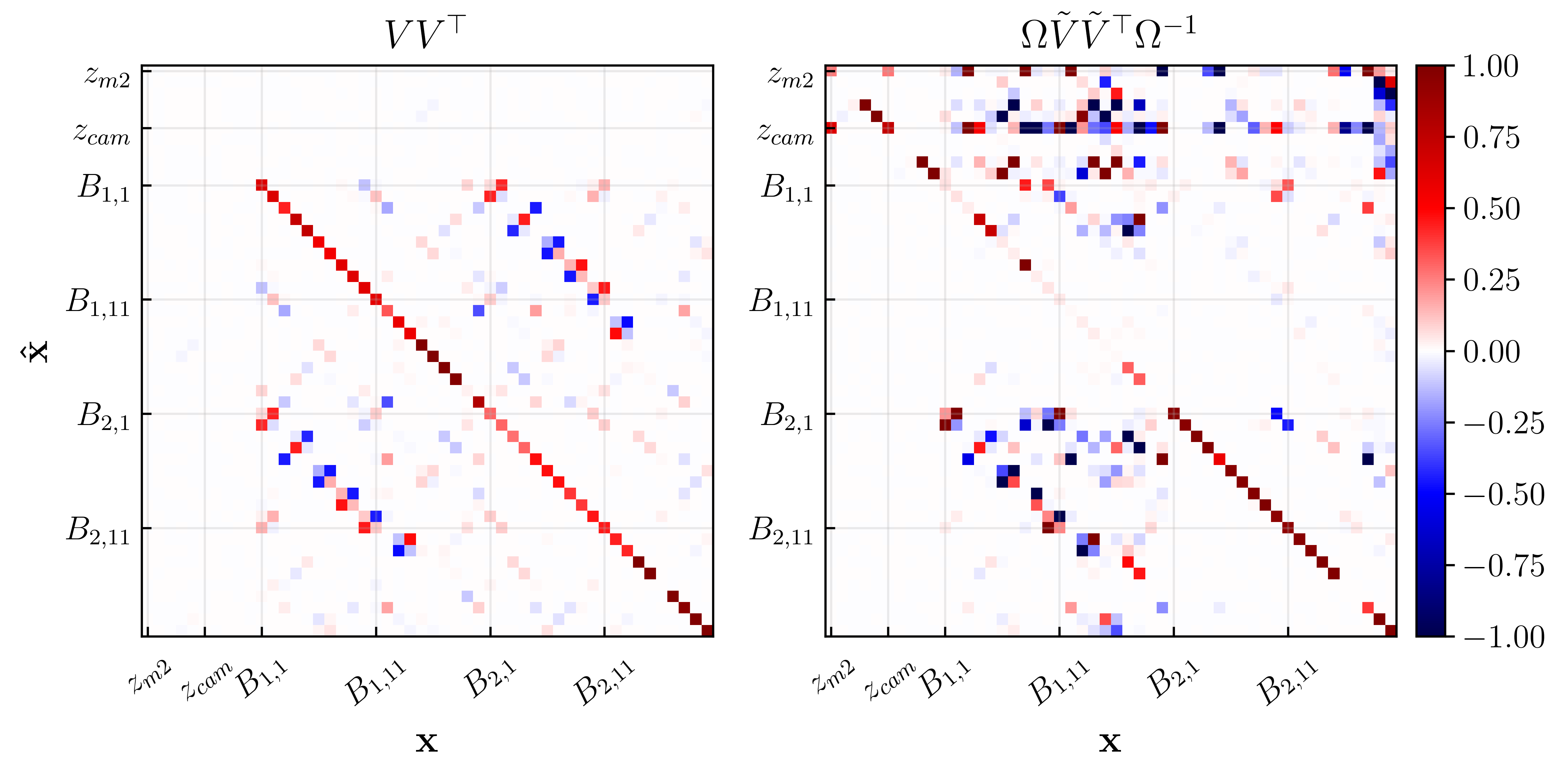}
    \caption{(Left) Displays the effective response matrix, $V V^{T}$, truncated at the 25th characteristic mode for the unscaled sensitivity matrix. (Right) Shows the effective response matrix, $\Omega \tilde{V} \tilde{V}^{\tr}\Omega^{-1}$, truncated at the 25th characteristic mode for the sensitivity matrix after applying the scaling factor $\Omega$. This side highlights how scaling influences the system's response, favoring M2 modes and including rigid body motions.}
    \label{fig:vvt}
\end{figure*}

\subsection{Noise effects}

The original sensitivity matrix, generated by displacing each degree of freedom by one micron \textemdash or one arcsecond for tips and tilts\textemdash, exhibits significant ill-conditioning, with a conditioning number of $\kappa(A) = 8 \cdot 10^5$. This underscores the need to exclude modes associated with minimal singular values. To assess the impact of noise and determine an appropriate truncation threshold, we introduce in Figure \ref{fig:power_spectrum} the power spectrum as defined in Equation \ref{power_eq}, applied to the rescaled basis. Accompanying this, we present the cumulative sum of singular values, setting the stage for our noise effect analysis and truncation criteria.

The left panel of Figure \ref{fig:power_spectrum} compares the power spectrum of the true wavefront deviation, derived from the Batoid optical model, against the estimated wavefront deviation calculated from simulated out-of-focus images captured by the four corner sensors (as described in Section \ref{sec:simulations}). The line plots average the results from the series of 100 distinct optical states, each simulated under different atmospheres with an average seeing of 0.8 arcsec of FWHM. Several features of these plots merit attention. The power spectrum from the true wavefront deviation, which excludes atmospheric effects, exhibits a steeper decline in the spectrum at modes with low singular values. Interpreted in terms of frequency, it becomes apparent that higher frequency components—those associated with lower singular values—contribute less significantly. In contrast, the estimated wavefront deviation shows an elevated contribution from high-frequency $\tilde{u}$ modes, underscoring the effect of noise in their estimation. This noise originates from both atmospheric conditions and algorithmic inaccuracies. Notably, algorithmic errors, as identified by \cite{Crenshaw_2024}, predominantly impact the Zernike estimates in Rubin Observatory's baseline TIE method for wavefront sensing. Indeed, our findings indicate that within reasonable ranges, the impact of different seeing conditions and star magnitude is minimal compared to the inaccuracies introduced by the wavefront estimation process. This observation led to the decision to simulate our out-of-focus images with 14th magnitude stars and with atmospheric conditions at 0.8 arcsec FWHM of seeing.

The addition of further wavefront error measurements across the field, for instance by rotating the camera, could allow us to better constrain the Zernike estimates and thereby minimize the power spectrum. Conversely, should there be fewer Zernike measurements\textemdash such as if no donut image falls within one of the corner wavefront sensors or if one ceases to function\textemdash our measurements would degrade, making it more challenging to constrain the degenerate space.

The manifestation of noise becomes pronounced from mode $\tilde{u}_{40}$ onwards for two primary reasons. Modes beyond $\tilde{u}_{50}$ are exclusively included because of SVD basis completion which extends the basis from 50 $u_i$ singular vectors up to 76 through the null space of the sensitivity matrix, they are thus absent in our measurements and bear null singular values. Modes between $\tilde{u}_{40}$ and $\tilde{u}_{50}$ correspond to those with the lowest singular values in Figure \ref{fig:weighted_svd}, with minimal impact on the wavefront. However, the onset of noise appears to be close to $\tilde{u}_{25}$, with the first noise peak being observed at $\tilde{u}_{30}$, which corresponds in Figure \ref{fig:weighted_svd} to a characteristic mode involving a quasi-degenerate counter-movement of pistoning between M2 and the Camera. 

This analysis, in conjunction with the right panel's cumulative singular value sum, guides our selection of the truncation threshold. The power spectrum suggests mode 25 as a plausible truncation point, a conclusion bolstered by the cumulative sum indicating that 99\% of the singular value contribution is reached by the 25th mode ($\tilde{\sigma}_{25}$).

Figure \ref{fig:vvt} presents the effective response matrix for both unscaled and scaled sensitivity matrices after truncation at this threshold. These matrices compare the true degrees of freedom on the $x$ axis, with the estimated degrees of freedom in the $y$ axis. More precisely, each column within these matrices indicates what each degree of freedom would be estimated to be, highlighting the characteristic modes discussed in the previous section. In the unscaled matrix, a distinct division between rigid body motions and bending modes is evident; without scaling, the truncation of the ten modes with the lowest singular values effectively excludes rigid body motions from all the estimates. Additionally, the unscaled matrix's lower diagonal block highlights the coupling between M1M3 and M2 modes assuming equal available range in both mirrors. In contrast, the scaled matrix adeptly reconstructs rigid body motions, including pistoning and tips/tilts, while excluding decenterings. The remainder of the scaled effective response matrix corroborates our previous observations: estimates predominantly interpret M1M3 bending modes as M2 modes alongside certain rigid body motions, while almost perfectly reconstructing M2 modes.

\subsection{Controller}
\begin{figure*}[h]
    \centering
    \figurenum{8}
    \epsscale{1.1}
    \plotone{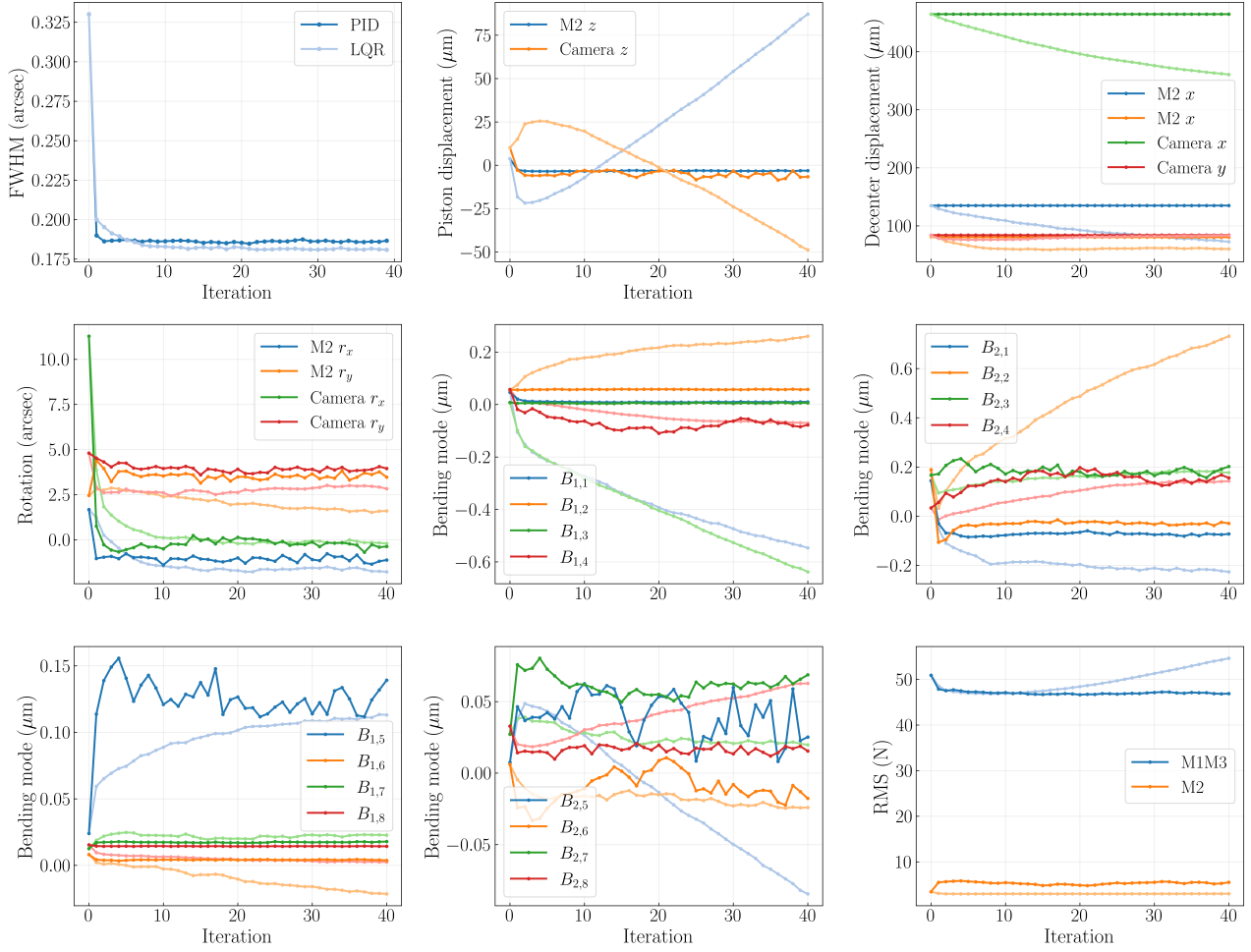}
    \caption{Progression of key performance indicators for OIC and PID controllers over multiple iterations, corresponding to a sequence of elevations from the LSST baseline survey. The top left graph highlights the FWHM evolution for both controllers. Surrounding plots detail the evolution across various degrees of freedom, distinguishing PID corrections (in a deeper color) from OIC corrections (in a lighter color). The bottom right graph focuses on the force actuator RMS evolution for the M1M3 and M2 mirrors, providing insight into the dynamic adjustments by both control strategies.}
    \label{fig:controller}
\end{figure*}

We implemented a PID controller using the estimated state from our reduced-order model, excluding the degenerate modes, and compared its performance to the existing OIC baseline. We tune the controller to determine $K_p$, $K_i$, and $K_d$ through the standard Ziegler-Nichols method \citep{Ziegler_1945}. With this method the gains are derived from an ``ultimate gain'', $K_u$, which we find to be 0.7. Figure \ref{fig:controller} presents the evolution of key metrics over multiple iterations for the OIC and PID approaches, simulating elevation changes during telescope operation following a sequence derived from the LSST baseline survey starting from a realistic optical state (as detailed in Section \ref{sec:simulations}). The performance metrics include the evolution of average Full Width at Half Maximum (FWHM) image quality averaged over the four corner wavefront sensors, the different system degrees of freedom, and the Root Mean Square (RMS) force exerted on the M1M3 and M2 actuators across iterative cycles. 

The OIC method achieves image quality convergence but fails to suppress movements within the noise-induced degenerate subspace. This limitation arises because the current OIC control strategy lacks a regularization term that effectively penalizes the movements in this subspace, since it penalizes the incremental motion instead of the total optical state. While future work will explore different penalty possibilities, these will still to be founded on the degenerate subspace that we have identified here.

The PID controller with our reduced-order basis demonstrates clear advantages. Firstly, compared to OIC, PID achieves faster image quality convergence. Nonetheless, we consider this advantage marginal given the tunable nature of PID gains and the expectation that initial iterations of the loop will have lower quality at the beginning of the night, as was found in DECam by \cite{Roodman_2014}. Additionally, PID reaches a slightly worse image quality of 0.005 arcsec, a minimal decrease within the FWHM resolution that is due to the added constraints in the parameter space with the reduced basis. Most importantly, however, is the PID ability to suppress degenerate mode movements. Compared to the OIC, Figure \ref{fig:controller} shows significantly reduced movement in all the degrees of freedom. There is still, however, a minor correlated oscillation between bending modes $B_{1,4}$ and $B_{2,4}$. Should this oscillation prove significant, further refinement of the degenerate basis could be warranted. Finally, PID maintains a consistent force actuator RMS over the iterations, unlike OIC, which exhibits a reduction followed by a steep increase linked to the divergence in the subspace of degeneracies.

These results highlight the effectiveness of our physically-informed, reduced-order model in improving telescope control. By excluding the degenerate subspace, the model enables the PID controller to achieve superior performance compared to the OIC approach.

\section{Discussion and Conclusion} \label{sec:conclusion}

Wide-field telescopes play a crucial role in astronomical surveys, yet their control systems face unique challenges due to the complex interplay between the degrees of freedom. Building on previous studies, notably the identification of a limited subspace of benign misalignments by \cite{Schechter_2011}, our investigation uncovers a larger \textit{subspace of noise-induced degeneracies}. Our findings demonstrate that eliminating these degeneracies simplifies the control strategy and improves overall system efficiency.

The emergence of this subspace arises from the coupling between mirror bending modes and misalignments, of which certain combinations have minimal impact on wavefront error. The presence of noise in wavefront estimates hinders our ability to determine the origin of the optical aberrations, rendering certain different combinations of degrees of freedom indistinguishable. Hence, discerning and prioritizing corrections among various telescope components becomes critical, with system constraints serving as key guidelines for these decisions.

To navigate this complexity, we propose a novel methodology for identifying the subspace of degeneracies, employing Singular Value Decomposition (SVD) on the system's sensitivity matrix, complemented by a rescaling approach. This technique adeptly captures the intricate relationships between degrees of freedom while accommodating their operational ranges and their impacts on the wavefront. We showcase the applicability of this method through its implementation at the Rubin Observatory, where our analysis decisively prioritizes the bending modes of M2 and hexapod movements over those of M1M3, thereby protecting the observatory's most critical mirror components.

The rescaling of the sensitivity matrix ensures the suppression of degenerate motions by embedding system-specific design constraints and priorities, resulting in optimized performance. Interestingly, our approach is adaptable and can be applied to various systems and different design constraints, yielding a different non-degenerate basis for each configuration.

To determine the degeneracy threshold, we present simulation-based noise study results comparing true and estimated wavefronts via the TIE approach, visualized through a power spectrum plot. While the true wavefront error may not be directly measurable in real-world, a simulated model remains valuable for comparison with noisy estimation values. This approach allows us to identify the noise threshold across multiple images.

Our results demonstrate that, as opposed to the OIC baseline, truncating the degenerate subspace during telescope control effectively prevents movements within it, leading to an optimal solution that minimizes telescope motion. The Vera C. Rubin Observatory serves as a pioneering application for this methodology and we plan for integration into the telescope's control code for testing during on-sky commissioning. This implementation will allow for real-world validation of our findings and potential performance improvements.

While this work primarily explores model reduction and state estimation, it opens the door to future explorations in advanced control strategies. We believe our methodology for identifying the subspace of degeneracies could be applied to the OIC baseline by informing an improved penalty term. Additionally, Model Predictive Control (MPC) is a promising potential avenue for ensuring the telescope stays within operational limits. While our closed-loop simulation suggests limited risk, longer simulations would be necessary to assess this risk and potentially implement safeguards. Such safeguards would not only prevent excursions into the noise-induced degenerate subspace but also ensure the telescope stays within its operational range.

Using the Vera C. Rubin Observatory as a practical example, our methodology and formulation provide a valuable framework for the next generation of large telescopes, including the European Southern Observatory's Extremely Large Telescope (ELT) and the Giant Magellan Telescope (GMT). While each telescope will have distinct system requirements and constraints, our approach readily accommodates these variations in the determination of the degenerate subspace.

In conclusion, this work addresses a critical aspect of telescope optics, paving the way for enhanced image quality in large-scale astronomical surveys, particularly for the Vera C. Rubin Observatory. This advancement will undoubtedly usher in a new era of scientific discoveries.

\begin{acknowledgements}
    We would like to thank Roberto Tighe and Aaron Roodman for helpful discussions on mirror prioritization, Tiago Ribeiro for discussions on the current baseline, Doug Niell and Ellie Hilerman for clarifications on the degree of freedom ranges and mirror flexures, Andy Rasmussen and Kevin Reil for reviewing this manuscript, and Will Sutherland for his optical intuition insights on an earlier version of this paper.

    This material is based upon work supported in part by the National Science Foundation through Cooperative Agreement AST-1258333 and Cooperative Support Agreement AST-1202910 managed by the Association of Universities for Research in Astronomy (AURA), and the Department of Energy under Contract No. DE-AC02-76SF00515 with the SLAC National Accelerator Laboratory managed by Stanford University. Additional Rubin Observatory funding comes from private donations, grants to universities, and in-kind support from LSST-DA Institutional Members. The project leading to these results also received funding from “la Caixa” Foundation (ID 100010434), under the agreement LCF/BQ/EU21/11890114. This research made use of Stanford University's computational resources.
    
    G.M. thanks the LSSTC Data Science Fellowship Program, which is funded by LSSTC, NSF Cybertraining Grant \#1829740, the Brinson Foundation, and the Moore Foundation; his participation in the program has benefited this work. J.F.C. acknowledges support from the U.S. Department of Energy, Office of Science, Office of High Energy Physics Cosmic Frontier Research program under Award Number DE-SC0011665.
\end{acknowledgements}

\vspace{5mm}

\facility{The Vera C. Rubin 8.4m Simonyi Survey Telescope Observatory}

\software{Batoid \citep{Meyers_2019},
          Galsim \citep{Galsim_2016}, 
          Matplotlib \citep{matplotlib_2007}, 
          NumPy \citep{numpy_2020}}

%

\vspace{5mm}


\bibliography{sample631}{}
\bibliographystyle{aasjournal}



\end{document}